\documentclass[10pt,preprint,a4paper]{aastex}
\usepackage{graphics,epsf}
\usepackage{amsmath}                % American Mathematical Society package
\usepackage{amsfonts}               % American Mathematical Society fonts
\usepackage{amssymb}                % American Mathematical Society symbol
\usepackage{epsfig}                 % EPS figures
\def \cm{~\rm{cm}}
\def \s{~\rm{s}}
\def \ms{~\rm{ms}}
\def \km{~\rm{km}}

\def \g{~\rm{g}}

\def \erg{~\rm{erg}}

\title{Jets launched at magnetar birth cannot be ignored}

\author{Noam Soker\altaffilmark{1}}

\altaffiltext{1} {Department of Physics, Technion -- Israel Institute of Technology, Haifa 32000 Israel; soker@physics.technion.ac.il}

\begin{document}

\begin{abstract}
I question models for powering super energetic supernovae (SESNe) with a magnetar central engine that do not include jets that are expected to be launched by the magnetar progenitor. I show that under reasonable assumptions the outflow that is expected during the formation of a magnetar can carry an amount of energy that does not fall much below, and even surpasses, the energy that is stored in the newly born spinning neutron star (NS).
The rapidly spinning NS and the strong magnetic fields attributed to magnetars require that the accreted mass onto the newly born NS possesses high specific angular momentum and strong magnetic fields. These ingredients are expected, as in many other astrophysical objects, to form collimated outflows/jets. I argue that the bipolar outflow in the pre-magnetar phase transfers a substantial amount of energy to the supernova (SN) ejecta, and it cannot be ignored in models that attribute SESNe to magnetars. I conclude that jets launched by accretion disks and accretion belts are more likely to power SESNe than magnetars are. This conclusion is compatible with the notion that jets might power all core collapse SNe (CCSNe).
\end{abstract}


% ==========================================================
\section{INTRODUCTION}
 \label{sec:intro}
% ==========================================================

Strongly-magnetized, rapidly-rotating neutron stars (NS), termed magnetars, have been suggested to power long-duration gamma-ray bursts and some super energetic supernovae (SESNe; e.g., \citealt{Metzgeretal2015} for a recent paper).
 These events are rare types of core-collapse supernova (CCSN) explosions of massive stars,
where more than $10^{53} \erg$ of gravitational energy is released during the formation of a NS or a black hole (BH) from the collapsing inner part of the progenitor's core. A small fraction of this energy is channelled to the kinetic energy of the ejected mass in the explosion.
There is as yet no consensus on the mechanism that channels a small fraction of the energy released in the collapse to the ejected mass.

The two contesting CCSN explosion mechanisms to utilize the gravitational energy are the delayed neutrino mechanism (e.g.,
\citealt{Bethe1985}; see review by \citealt{Janka2012})
and explosion based on jets launched by the newly formed NS or BH and operating via a jet feedback mechanism  (JFM; e.g., \citealt{Papishetal2016} and \citealt{Gilkisetal2016}, and references therein). Another explosion mechanism is the collapse-induced thermonuclear explosion (CITE) mechanism \citep{Burbidgeetal1957, KushnirKatz2014}, but it attributes the energy to thermonuclear burning rather than to gravitational energy.

By SESNe I refer to CCSNe with explosion (kinetic plus radiation) energy above about $10^{52} \erg$, e.g., SN~2010ay with a kinetic energy of $E_\mathrm{kin} \approx 1.1 \times 10^{52} \erg$ \citep{Sandersetal2012}.
The question of the powering of SESNe is a hot topic (e.g., \citealt{Quimbyetal2013, Moriyaetal2015, Wangetal2016, Arcavietal2016, Sorokinaetal2016}; see review by \citealt{GalYam2012}).
The JFM can account for SESNe with the same basic JFM that works for regular CCSNe, but instead of jittering-jets that work for most regular CCSNe \citep{Papish2011}, the jets have a more or less constant axis \citep{Gilkisetal2016}. The delayed neutrino mechanism on the other hand, is not intended to account for such energies (e.g., \citealt{Papishetal2015}), and an extra energy source is required. Such are the idea of quark-nova (e.g., \citealt{Ouyedetaletal2016}) and suggestions of extra powering by a magnetar (e.g., \citealt{KasenBildsten2010, Woosley2010, Metzgeretal2015}).
The recent SESN ASASSN-15lh with an explosion energy of $> 1.1 \times 10^{52} \erg$ \citep{Dongetal2016} challenged the magnetar mechanism (e.g., \citealt{Daietal2016, Metzgeretal2015, Berstenetal2016}). \cite{Metzgeretal2015} reexamine the magnetar mechanism and claim that it can indeed account for the extra energy required to explain the SESN ASASSN-15lh.

In this short letter I argue that the formation process of such magnetars is very likely to be accompanied by the launching of jets that carry a substantial amount of energy, and hence cannot be ignored. Numerical simulations have shown that when a rapidly rotating core with ordered magnetic fields collapses to form a rapidly rotating NS, jets are launched, e.g., \cite{Nishimura2015}. I do not assume ordered magnetic fields, but rather that they can be amplified by a dynamo in the accretion flow close to the NS (sec. \ref{sec:AM}). In section \ref{sec:energy} I consider the possible energy that is carried by such jets. I summarize this short study in section \ref{sec:summary}.

% ==========================================================
\section{SPECIFIC ANGULAR MOMENTUM}
 \label{sec:AM}
% ==========================================================

One first must distinguish between the baryonic mass that forms the NS and the final mass of the NS. The latter is smaller, as neutrinos carry energy equivalent to a mass $\Delta M_{\nu}$ from the newly born NS. Neutrinos carry away angular momentum $\Delta J_{\nu}$ as well \citep{Camelioetal2016}. The neutrinos are scattered from the neutrinosphere, such that they leave the system with an average tangential velocity about equal to that of the gas in the neutrinosphere. Therefore, the specific angular momentum of the neutrinos, $j_{\nu} \equiv \Delta J_{\nu} / \Delta M_{\nu}$, is about equal to that of the accreted mass at the neutrinosphere at a given moment.
This approximate analysis shows that the specific angular momentum of the accreted baryonic mass is about equal to that of NS if angular momentum is not removed by an outflow.

Let us start with the assumption that angular momentum is not removed by an outflow of gas, and let us follow \cite{Metzgeretal2015} and take the following properties of a magnetar at maximum angular velocity possible. Other properties of the NS are based on \cite{LattimerSchutz2005}.
The moment of inertia of the NS is $I_{\rm NS} \approx 1.3 \times 10^{45} (M_{\rm NS}/1.4 M_\odot)^{3/2} \g \cm^2$. \cite{Metzgeretal2015} take the minimum period for a NS of mass $M_{\rm NS}=2 M_\odot$ to be $P_m=0.7 \ms$, such that the rotational energy is $E_{\rm NS-max} (2 M_\odot) \approx 9 \times 10^{52} \erg$, and its angular momentum is $J_{2.0} \approx 2.0 \times 10^{49} \erg \s$.

I consider the first few seconds where the mass of the last $0.6 M_\odot$ is accreted onto the NS.
This is the final mass, the baryonic mass that is accreted is larger even. The outer part of the NS is still at a larger radius than its final radius, but not by much. In any case, the mass layers must contract continuously toward the newly born NS.
I further assume that when the NS mass is $1.4M_\odot$ it rotates at its maximum possible rate, with a period of $P \approx 1 \ms$. The angular momentum of the NS is
$J_{1.4} \approx 8.2 \times 10^{48} \erg \s$.
Therefore, the last $0.6 M_\odot$ is accreted with a total angular momentum of
$\Delta J_{\rm acc}  = J_{2.0} - J_{1.4} \simeq  1.2 \times 10^{49} \erg \s$, and a specific angular momentum of $j_{\rm acc} \simeq 1.0 \times  10^{16} \cm^2 \s^{-1}$.
The specific angular momentum of a Keplerian motion at $r= 15 \km$ from the centre of the final NS of mass $2 M_\odot$ is $j_{\rm Kep} \simeq 2 \times 10^{16} \cm^2 \s^{-1}$. This implies
\begin{equation}
\label{eq:j1}
\frac{j_{\rm acc}}{j_{\rm Kep}}
\simeq 0.5
\left( \frac{\Delta J_{\rm acc}}{1.2 \times 10^{49} \erg \s}  \right)
\left( \frac{\Delta M_{\rm acc}}{0.6 M_\odot}  \right)^{-1}
\left( \frac{r}{15 \km}  \right)^{-1/2}
\cm^2 \s^{-1}.
\end{equation}

Mass is accreted not only from the equatorial plane. If mass at higher latitudes has lower specific angular momentum than the mass near the equatorial plane, then the mass near the equatorial plane has a larger value of
 ${j_{\rm acc}}/{j_{\rm Kep}}$ than the average value given in equation (\ref{eq:j1}).
Although a Keplerian accretion disk is not formed, at least not immediately, an accretion belt of gas is formed around the newly formed NS.
Such an equatorial accretion belt develops a strong magnetic activity that is very likely to launch a bipolar outflow \citep{SchreierSoker2016}. {{{{ The formation of jets in a CCSN with a rapidly rotating core that is likely to lead to the formation of a magnetar has been simulated by  \cite{Mostaetal2015}. In their simulations an accretion disk was formed. Here I claim that even a sub-Keplerian outflow can lead to jet formation.   }}}}

% ==========================================================
\section{THE ENERGY OF THE BIPOLAR OUTFLOW}
 \label{sec:energy}
% ==========================================================

In the derivation of equation (\ref{eq:j1}) it was assumed that the NS is rotating at the maximum speed possible. The accreted mass has about half the maximum possible rotation velocity (depends on the latitude). A very strong shear exists between the accreted mass and the NS. This will lead to turbulence, that together with the velocity shear leads to the amplification of the magnetic fields. The magnetic activity in turn might launch bipolar outflow through the polar regions \citep{SchreierSoker2016}.
The study of this flow requires a sophisticated 3D magneto-hydrodynamical code. Here I limit the discussion to the estimation of the possible energy of the outflow.

The velocity difference between the NS and the accreted mass in the equatorial plane derived from equation (\ref{eq:j1}) is  $\Delta v \approx 6.7 \times 10^4 \km \s^{-1}$. The kinetic energy to be dissipated due to this velocity difference is
 \begin{equation}
E_{\rm diss} \approx  \frac{1}{2} M_{\rm acc} (\Delta v )^2
\approx 2 \times 10^{52}
\left( \frac{M_{\rm acc}}{0.6 M_\odot}  \right)
\left( \frac{\Delta v}{6 \times 10^4 \km \s^{-1}}  \right)^{2}
\erg.
\label{eq:e1}
\end{equation}
We can scale quantities in a different way. Based on other astrophysical objects, such as young stellar objects (e.g. \citealt{VorobyovBasu2010}), we take $10 \%$ of the accreted mass, $\approx 0.06 M_\odot$, to be launched into jets at the escape speed from the NS, $v_{\rm esc} \simeq 1.5 \times 10^5 \km \s^{-1}$. This amounts to a kinetic energy of
 $E_{\rm jets} \approx 10^{52} \erg$.

Over all, the above discussion raises the possibility that the formation of a magnetar with an energy of $E_{\rm NS-max} (2 M_\odot) \approx 9 \times 10^{52} \erg$ is likely to result in the launching of jets with a total energy of $\approx 1-2 \times 10^{52} \erg$. This is an order of magnitude higher than the typical explosion (kinetic) energy of typical CCSNe.
Moreover, the expected jets carry some angular momentum as well. This implies that with the above initial conditions, the final energy of the magnetar will be smaller than the maximum allowed value $E_{\rm NS} < E_{\rm NS-max}$.

% ==========================================================
\section{SUMMARY}
 \label{sec:summary}
% ==========================================================

In this short study I argued that jets are very likely to be launched by the newly born NS as it accretes mass to form an energetic magnetar.  In section \ref{sec:AM} I showed that the formation of a magnetar of $2 M_\odot$ with the maximum possible energy, requires that the last $\approx 0.6 M_\odot$ be accreted with very high specific angular momentum. Although a Keplerian disk is not necessarily formed, an equatorial accretion belt is formed. A large velocity shear exists between the accreted mass and the NS. This shear is expected to lead to the development of turbulence and to substantial amplification of the magnetic fields \citep{SchreierSoker2016}. The magnetic activity, such as reconnection and line-winding, might launch jets.
A similar idea was studied by \cite{Akiyamaetal2003} and \cite{Mostaetal2015}, who argued that the  magnetorotational instability (MRI) and dynamo lead to a bipolar outflow.

In section \ref{sec:energy} I argued that if jets are launched, they can carry an energy of
$E_{\rm jets} \simeq 1-2 \times 10^{52} \erg$. This is much more than a typical explosion energy of CCSNe. This energy is smaller than the maximum energy of the magnetar. But if jets are indeed launched, they are expected to carry away some of the angular momentum, and hence the magnetar will not be at its maximum possible energy.

Jets with a well defined axis are less efficient in ejecting the mass from regions close to the equatorial plane. As more mass is accreted to the center, a BH is formed
\citep{Gilkisetal2016}. In this jet feedback mechanism (JFM) scenario a BH is formed not from a failed CCSN, but rather from a more violent CCSN. It is more violent in the sense that the total kinetic energy is larger than when a NS is formed, and the BH is likely to launch relativistic jets. The explosion energy is determined by the efficiency of the JFM \citep{Gilkisetal2016}, and not by neutrino emission or magnetar activity. Generally, a less efficient feedback will lead to a more energetic explosion! Models that attribute the most energetic CCSNe to the operation of a magnetar and ignore jets (e.g., \citealt{SukhboldWoosley2016} for a most recent study) are not complete.

Although I did not show that jets must be launched during the formation of a magnetar, the present study does question magnetar-powered models for SESNe that ignore jets launched during the formation process of the magnetar. \emph{Models for powering SESNe with a magnetar central engine must justify not including jets that are expected to be launched by the magnetar progenitor. }

\vspace*{0.5cm}
%\section*{Acknowledgments}

{}

\end{document}